\documentclass{article}
\input epsf

\begin{document}

\title{Spin density matrices for nuclear density functionals with parity 
violations}

\author{B. R. Barrett \\
 Physics Department, University of Arizona, \\
Tucson, AZ 85721, USA \\
        B. G. Giraud \\
Institut de Physique Th\'eorique, DSM, CE Saclay, \\
91191 Gif-sur-Yvette, France}      

\date{\today} 
\maketitle

\begin{abstract}

The spin density matrix (SDM) used in atomic and molecular physics is 
revisited for nuclear physics, in the context of the radial density functional 
theory. The vector part of the SDM defines a ``hedgehog'' situation,
which exists only if nuclear states contain some amount of parity violation.

\end{abstract}

\begin{center}
\end{center}

\section{Introduction}
The subject of density functionals (DFs) in nuclear physics \cite{unedf, DFP} 
is presently receiving intense attention. One of its difficulties is the 
handling of interactions that depend on spins. While there is {\it a priori} 
no theorem preventing a theory with simple densities from accommodating the 
influence of spin dependent forces, it is likely that a generalization of 
density profiles to ``spin density'' ones should, in practice, make the 
construction of a DF easier. The purpose of this paper is see whether one
can adapt to nuclear physics the same concept \cite{GunLun, Goe} as 
that used for many years in atomic and molecular physics. 

Given the creation and annihilation operators, $a_{\vec r \sigma}^{\dagger}$ 
and $a_{\vec r \sigma},$ respectively, of a nucleon with spin 
$\sigma=\pm \frac{1}{2}$ at position $\vec r,$ and given a density operator 
${\cal D}$ in many-body space, the SDM, $\bar \rho(\vec r),$ is defined by 
its matrix elements,
\begin{equation}
\rho_{\sigma \sigma'}(\vec r)=
{\rm Tr}\, a_{\vec r\, \sigma}^{\dagger}\,  a_{\vec r\, \sigma'}\, {\cal D}
\, . 
\end{equation}
In the following, we shall take advantage of the recent proof \cite{Gir}, 
based upon the rotational invariance of the nuclear Hamiltonian, that the 
nuclear DF is a scalar, namely a radial density functional (RDF); accordingly,
it is understood in the following, unless explicitly stated otherwise, that 
the density operator, ${\cal D},$ in many-body space, is a scalar under 
rotations. Since practical calculations for a DF can eventually result in 
Kohn-Sham (KS) potentials \cite{KS}, the approach described by the present 
paper, with its explicit treatment of spin, might give indications for the 
spin-orbit term in KS equations.

The basic formalism for the SDM is explained in Sec. 2. A mandatory
generalisation of the formalism is explained in Sec. 3. An illustrative 
example is provided in Sec. 4. We conclude in Sec. 5.

\section{Basic formalism}
We first relate the local creation and annihilation operators to those of an 
$\ell s$ shell model, 
\begin{equation}
a_{\vec r\, \sigma }^{\dagger}=\sum_{n \ell m} \varphi_{n \ell }(r)\, 
Y_{\ell m }^*(\hat r)\, a_{n \ell m \sigma }^{\dagger}\, ,\ \ \ \ 
a_{\vec r\, \sigma'}          =\sum_{n' \ell' m'} \varphi_{n' \ell' }(r)\, 
Y_{\ell' m'}  (\hat r)\, a_{n' \ell' m'\sigma'}          \, .
\label{lsscheme}
\end{equation}
Here the wave functions, $\varphi_{n \ell}(r)\, Y_{\ell m}(\hat r),$ represent 
the orbitals created by the operators $a_{n \ell m}^{\dagger},$  with real 
radial form factors $\varphi_{n \ell}(r).$ The summation, $\sum_{n \ell m},$ 
runs over a complete basis of orbitals, assumed to be discrete for the sake of 
simplicity. A generalization with continuum orbitals brings no difficulty 
except for slightly less simple notations. Isospin labels are understood.

We then rearrange the products, 
$a_{\vec r \sigma}^{\dagger} a_{\vec r \sigma'},$ into their scalar and 
vector parts in spin space with the usual Clebsch-Gordan 
coefficients,
\begin{equation}
A_{\vec r S M_S}=\sum_{\sigma \sigma'} (-)^{\frac{1}{2}-\sigma'}
\langle \frac{1}{2}\, \sigma\, \frac{1}{2}\, -\sigma' | S\, M_S \rangle\,
a_{\vec r \sigma}^{\dagger} a_{\vec r \sigma'},
\label{spncoupl}
\end{equation}
This gives, after inserting Eqs. (\ref{lsscheme}),
\begin{equation}
A_{\vec r S M_S}=
\sum_{ \ell m \ell' m'} Y_{ \ell m}^*(\hat r)\, Y_{ \ell' m'}(\hat r)\,
B_{ \ell m \ell' m' S M_S}(r)\, ,
\end{equation}
with
\begin{eqnarray}
&&B_{\ell m \ell' m' S M_S}(r)= \nonumber \\
&&\sum_{n n' \sigma \sigma'} 
\varphi_{n \ell}(r)\, \varphi_{n' \ell'}(r)
\, (-)^{\frac{1}{2}-\sigma'}
\langle \frac{1}{2}\, \sigma\, \frac{1}{2}\, -\sigma' | S\, M_S \rangle\,
a_{n \ell m \sigma}^{\dagger} a_{n' \ell' m' \sigma'}\, .
\label{prncpsum}
\end{eqnarray}
Next we recouple the orbital momenta carried by the operators 
$B_{\ell m \ell' m' S M_S}\, ,$
\begin{equation}
C_{\ell \ell' L M S M_S}(r) = \sum_{m_1 m_2} (-)^{\ell'-m_2}\,
\langle \ell\, m_1\, \ell'\, -m_2 | L\, M \rangle\,
B_{\ell m_1 \ell' m_2 S M_S}(r)\, ,
\label{orbtcoupl}
\end{equation}
so that
\begin{equation}
A_{\vec r S M_S}= 
\sum_{\ell m \ell' m'} Y_{\ell m}^*(\hat r)\, Y_{\ell' m'}(\hat r)\, 
(-)^{\ell'-m'}\, \sum_{L M} \langle \ell\, m\, \ell'\, -m' | L M \rangle\, 
C_{\ell \ell' L M S M_S}(r)
\, .
\label{decoupl}
\end{equation}
Upon taking advantage of the relations i) between spherical harmonics,
\begin{eqnarray}
Y_{\ell m}(\hat r)\, Y_{\ell' m'}^*(\hat r) &=&  (-)^{m'}\, 
\sum_{\lambda \mu} 
\sqrt{\frac{(2 \ell +1)\, (2 \ell' +1)\, (2 \lambda +1)}{4 \pi}}\ \times
\nonumber \\
&& \left(\matrix{\ell & \ell' & \lambda \cr 0 &  0  &  0  }\right)\, 
   \left(\matrix{\ell & \ell' & \lambda \cr m & -m' & \mu }\right) 
Y_{\lambda \mu}^*(\hat r)\, ,
\end{eqnarray}
and ii) between Wigner $3j$-coefficients and Clebsch-Gordan ones,
\begin{equation}
\langle l\, m\, l'\, -m' | L\, M \rangle = (-)^{\ell-\ell'+M}\, \sqrt{2L+1}\,
\left(\matrix{\ell & \ell' & L \cr m & -m' & -M }\right)\, ,
\end{equation}
the orthogonality between Wigner $3j$-coefficients,
\begin{equation}
\sum_{mm'} (2L+1)\,
\left(\matrix{\ell & \ell' &    L    \cr m & -m' & -M  }\right)\, 
\left(\matrix{\ell & \ell' & \lambda \cr m & -m' & \mu }\right) = 
\delta_{L \lambda}\, \delta_{-M\, \mu}\, ,
\end{equation}
simplifies Eq. (\ref{decoupl}) into,
\begin{eqnarray}
A_{\vec r S M_S} &=& \sum_{\ell \ell' L M} (-)^{L-M}\, Y_{L-M}(\hat r) \times
\nonumber \\
&&(-)^{\ell'}\, \sqrt{\frac{(2 \ell +1)(2 \ell' +1)}{4\pi}}\,
\left(\matrix{\ell & \ell' & L \cr 0 & 0 & 0} \right)\, 
C_{\ell \ell' L M S M_S}(r)\, .
\label{simpl}
\end{eqnarray}
In Eq. (\ref{simpl}), we used the facts that all numbers, 
$\ell,\ell',L,M,$ are integers and that the $3j$-coefficient, 
$\left(\matrix{\ell & \ell' & L \cr 0 & 0 & 0}\right),$ vanishes unless
$\ell+\ell'+L$ is even. 

Finally, a recoupling of total orbital momentum and total spin yields,
\begin{eqnarray}
&&D_{L S J\mu}(r)=\sum_{M M_S} \langle L M S M_S | J\mu \rangle\ \times 
\nonumber \\ 
&&\left[ \sum_{\ell \ell'}\, (-)^{\ell'} 
\sqrt{\frac{(2 \ell+1) (2 \ell'+1)}{4 \pi}}
\left(\matrix{\ell & \ell' & L \cr 0 & 0 & 0}\right)
C_{\ell\ell' L M S M_S}(r) \right] ,
\label{totcoupl}
\end{eqnarray}
so that,
\begin{eqnarray}
&&\sum_{\ell \ell'}\, (-)^{\ell'}\, 
\sqrt{\frac{(2 \ell+1) (2 \ell'+1)}{4 \pi}}\, 
\left(\matrix{\ell & \ell' & L \cr 0 & 0 & 0}\right)\, 
C_{\ell \ell' L M S M_S}(r)
= \nonumber \\
&&\sum_{J \mu} \langle L M S M_S | J\mu \rangle\, D_{LSJ\mu}(r)\, .
\end{eqnarray}
In terms of $D_{LSJ\mu}(r),$ the scalar or vector operators for the SDM 
now read
\begin{equation}
A_{\vec r S M_S}= \sum_{L M} (-)^L\, Y_{L M}^*(\hat r) \sum_{J \mu}
\langle L\, M\, S\, M_S | J \mu \rangle\, D_{L S J \mu}(r)\, .
\label{expansion}
\end{equation}
With scalar density matrices ${\cal D}$ in many-body space, there
will be vanishing traces, ${\rm Tr}\, D_{L S J\mu}(r)\, {\cal D},$ 
unless $J=\mu=0.$ In this case, the corresponding Clebsch-Gordan coefficient 
becomes,
\begin{equation}
\langle L\, M\, S\, M_S |  0\, 0 \rangle = \delta_{LS}\ \delta_{M\, -M_S}\
\frac{ (-)^{S+M_S} }{ \sqrt{2S+1} }\, ,
\end{equation} 
so that the SDM scalar or vector elements reduce to,
\begin{equation}
{\rm Tr}\, A_{\vec r S M_S}\, {\cal D}=\frac{(-)^{M_S} \, 
Y_{S\, -M_S}^*(\hat r)}{\sqrt{2S+1}}\, {\rm Tr}\, D_{S S 0 0}(r)\, {\cal D}
= \frac{Y_{S M_S}(\hat r)}{\sqrt{2S+1}}
\, {\rm Tr}\, D_{SS00}(r)\, {\cal D}\,
\label{compact} .
\end{equation}

\section{Generalization}
Two very different spin profiles emerge from the study made in Sec. 2. For the 
first of them, namely, for $S=0,$ the result is simple, since, necessarily  
in this case, $\ell$ and $\ell'$ are equal,
\begin{equation}
D_{0000}(r)= \frac{1}{ \sqrt{ 8 \pi } }\, \sum_{n n'\ell m \sigma}
\varphi_{n \ell}(r)\, \varphi_{n' \ell}(r)\, 
a_{n \ell m \sigma}^{\dagger}\, a_{n'\ell m \sigma}\, .
\end{equation}
For the second profile, {\it i.e.,} for $S=1,$ spherical symmetry 
is ensured by the fact that all three spherical harmonics are multiplied by 
the same, radial form factor, which we denote $\rho_{hh}(r)$ in the 
following; we have a `hedgehog'' situation. 
Here we mean hedgehog-like in the sense that the vector spin field 
has only a radial dependency.
It must be noticed, however, that only those pairs of particle orbital 
momenta $\{\ell,\ell'\},$ where $|\ell-\ell'|\leq 1,$ can couple to $L=1.$ 
If $\ell=\ell',$ the $3j$-coefficient,
$\left(\matrix{\ell & \ell' & L \cr 0 & 0 & 0}\right),$ vanishes identically, 
since $\ell+\ell'+1$ becomes odd. Conversely, if $\ell-\ell'=\pm 1,$ the 
corresponding products of operators, 
$a_{n \ell m}^{\dagger} a_{n' \ell' m'},$ have an odd parity. Since parity 
violations in nuclear states are most often too tiny to be observable, the 
density operators ${\cal D}$ of interest always have an even parity. 
Therefore, if the traces, 
${\rm Tr}\, C_{\ell\, \ell\pm 1\, 1\, -M_S\, 1\, M_S}(r)\, {\cal D},$ 
do not vanish completely, then they will detect parity violations in 
${\cal D}.$ A basic RDF, that uses $D_{0000}$ only, has no easy signature for 
parity violations. It is the occurrence of a tiny, but non-vanishing profile 
from $D_{1100}$ that allows a more elaborate RDF theory to explicitly 
accommodate parity violations.

For the sake of completeness, we show in Eq. (\ref{hdg}) this ``hedgehog'' 
operator, $D_{hh}(r) \equiv D_{1100}/\sqrt{3},$ the trace of which with 
${\cal D}$ is the coefficient of $Y_{1 M_S}(\hat r)$ in Eq. (\ref{compact}). 
It reads, upon taking advantage of 
Eqs. (\ref{prncpsum}), (\ref{orbtcoupl}), (\ref{totcoupl}) and (\ref{compact}),
\begin{eqnarray}
D_{hh}(r)&=&
\frac{1}{3 \sqrt{4 \pi}}\, \sum_{n n'\ell m \ell' m' \sigma \sigma' M_S}\,
\varphi_{n \ell}(r)\, \varphi_{n' \ell'}(r)
\, (-)^{1+M_S-m'+\frac{1}{2}-\sigma'}\ \times  \nonumber \\
&& \sqrt{(2 \ell+1) (2 \ell'+1)}\, 
\left(\matrix{\ell & \ell' & 1 \cr 0 & 0 & 0}\right)\,
\langle \ell\, m\, \ell'\, -m' | 1\, -M_S \rangle\, \ \times \nonumber \\
&& \langle \frac{1}{2}\, \sigma\, \frac{1}{2}\, -\sigma' | 
1\, M_S \rangle\, a_{n \ell m \sigma}^{\dagger}\, a_{n'\ell' m' \sigma'}\, .
\label{hdg}
\end{eqnarray}

A natural way to enlarge the theory to cases where the $S=1$ form factor is 
not tiny consists in embedding the nucleus in an external field, $H_1,$ 
that simultaneously breaks the rotational symmetry and the parity. To 
avoid loosing the advantage of an RDF, {\it i.e.,} the reduction of 
three-dimensional calculations to one-dimensional ones, the symmetry breaking 
can be chosen as a minimal one, in the following way. Let $H_1$ be a negative 
parity operator, bounded from below, that transforms as a vector under 
rotations. There is no need to assume that $H_1$ is only made of local fields, 
$H_1=\sum_i h_1(\vec r_i,\sigma_i),$ where $\vec r_i$ and $\sigma_i$ denote 
the position and spin of the $i$th nucleon; any complicated $H_1$ is allowed 
for the argument to come. What counts is that the extended Hamiltonian,
$H'=H+H_1,$ which is bounded from below, now contains, besides the basic 
scalar and positive parity $H,$ a vector and negative parity component $H_1.$
Then we use the ``constrained search'' definition \cite{LevLie} of a DF,
\begin{equation}
F[\bar \rho] =
{\rm Inf}_{{\cal D}\rightarrow \bar \rho}\, {\rm Tr}\, H'\, {\cal D},
\end{equation}
where now ${\cal D}$ is generalized into an arbitrary density operator, 
without symmetry properties. Here the symbol, ${\cal D}\rightarrow \bar \rho,$
means that the minimization of the energy is performed over subsets in 
the ${\cal D}$ space that show a given spin density matrix $\bar \rho.$ Then
the same argument, as that used in \cite{Gir}, to restrict ${\cal D}$ to be 
a rotation scalar, can be extended to restrict ${\cal D}$ to be a mixture 
${\cal D}_{01}$ of a scalar and a vector. Next one can take advantage of 
Eq. (\ref{expansion}) and derive $F[\bar \rho]$ from those {\it few}
and {\it radial} profile operators, $D_{LSJ\mu}(r)$, where the conditions, 
$S=0,1$ and $J=0,1$ give limits to $L$ via the usual triangular rules.

To conclude this Sec., we note that a spin density DF is usually not very 
useful for an isolated nucleus, but becomes legitimate for a non-isolated one.

\section{Toy model for an illustrative example}
Consider a fictitious  $^{16}$O nucleus made of a full $0s$ shell and an 
almost full $0p$ shell and driven by a harmonic oscillator Hamiltonian, 
$$
H_0=\sum_{nlm \sigma \tau} \left( 2n+\ell+\frac{3}{2} \right) 
a_{n \ell m \sigma \tau}^{\dagger}\, a_{n\ell m \sigma \tau} =
\sum_{nlj \mu \tau} \left( 2n+\ell+\frac{3}{2} \right) 
b_{n \ell j \mu \tau}^{\dagger}\, b_{n \ell j \mu \tau}\, . 
$$
Here, temporarily, the isospin label, $\tau=\pm \frac{1}{2},$ is explicit. 
The relation between $\ell s$ and $jj$ creation operators (and, similarly, 
for annihilation ones) in this toy model reads, 
\begin{equation}
b_{n \ell j \mu    \tau}^{\dagger}=\sum_{m \sigma} \langle \ell m \frac{1}{2}
\sigma | j \mu \rangle\, a_{n \ell m \sigma \tau}^{\dagger}\, ,\ \ 
a_{n \ell m \sigma \tau}^{\dagger}=\sum_{j    \mu} \langle \ell m \frac{1}{2}
\sigma | j \mu \rangle\, b_{n \ell j \mu    \tau}^{\dagger}          \, . 
\end{equation}
A Slater determinant, $| \phi \rangle,$ will describe this nucleus for our 
model. Assume that a perturbation of the harmonic oscillator slightly 
mixes the $0p_{\frac{1}{2}}$ orbitals with the $1s_{\frac{1}{2}}$ orbitals. 
The mixtures read,
\begin{equation}
\beta_{\, j=\frac{1}{2}, \mu,\tau}^{\dagger} = \cos \varepsilon\ 
    b_{\, 0p\frac{1}{2},\mu,\tau}^{\dagger} + \sin \varepsilon\  
    b_{\, 1s\frac{1}{2},\mu,\tau}^{\dagger}\, .
\end{equation}
We keep intact a core, made of the $0s$ and $0p\frac{3}{2}$ orbitals.
The $16$-body operator, ${\cal D}=| \phi \rangle \langle \phi|,$ is still a 
scalar under rotations, but it has now a negative parity component at first
order in $\varepsilon.$  Such a state, and similar density matrices, would 
justify the use of a ``spin RDF'', with two profiles.

Let $|0\rangle$ denote the fully closed $0s$ and $0p$ shells. At first order
in $\varepsilon,$ the wave function under consideration is, 
$|\phi\rangle=|0\rangle+\varepsilon \sum_{\tau} |\tau\rangle,$ with
\begin{equation}
|\tau \rangle = \left(
b_{\, 1s\frac{1}{2}, \frac{1}{2}, \tau}^{\dagger}\, b_{\, 0p\frac{1}{2}, 
 \frac{1}{2}, \tau} + 
b_{\, 1s\frac{1}{2},-\frac{1}{2}, \tau}^{\dagger}\, b_{\, 0p\frac{1}{2}, 
-\frac{1}{2}, \tau} 
\right) | 0 \rangle .
\label{pathol}
\end{equation}
Protons and neutrons will give equal matrix elements; hence, within an 
inessential factor of 2, isospin labels and summations can again be omitted. 
Notice also, incidentally, that the particle-hole states $| \tau \rangle,$ 
shown in Eq. (\ref{pathol}), do not represent center-of-mass spurious shifts; 
the latter induce dipoles, not monopoles, in the one-particle-one-hole space.

In the $jj$ representation, we obtain for the $S = 0$ case
\begin{equation}
D_{0000}(r)= \frac{1}{ \sqrt{ 8 \pi } }\, \sum_{n n'\ell j}
\varphi_{n \ell}(r)\, \varphi_{n' \ell}(r)\, 
\sum_{\mu} b_{n \ell j \mu}^{\dagger}\, b_{n'\ell j \mu}\, ,
\label{D0000}
\end{equation}
and the scalar profile, ${\rm Tr} D_{0000} | \phi \rangle \langle \phi |,$
has a vanishing contribution from the first-order matrix elements, 
$\langle 0 | D_{0000} | \tau \rangle=\langle \tau | D_{0000} | 0 \rangle = 0,$
because of the restriction to equal values of $\ell$.
The zeroth-order profile from the $0s$- and $0p$-shells, respectively,
is obviously
\begin{equation}
D_{0000}(r) \propto \varphi_{00}^2(r)+3\, \varphi_{01}^2(r) =
\left( 2 \pi^{-\frac{1}{4}}\, e^{-\frac{1}{2}r^2}\right)^2 +
3 \left( 2^{\frac{3}{2}} 3^{-\frac{1}{2}} \pi^{-\frac{1}{4}}\, r\, 
e^{-\frac{1}{2}r^2}\right)^2,
\end{equation} 
with an inessential coefficient, $\sqrt{2/\pi},$ omitted for simplicity.

Again for the $jj$ representation, we find for the $S = 1$ (hedgehog) case,
\begin{eqnarray}
D_{hh}(r)&=&
\frac{1}{3 \sqrt{4 \pi}}\, 
\sum_{n n'\ell \ell' j j' \mu \mu' m  m' \sigma \sigma' M_S}\,
\varphi_{n \ell}(r)\, \varphi_{n' \ell'}(r)
\, (-)^{1+M_S-m'+\frac{1}{2}-\sigma'}\ \times  \nonumber \\
&& \sqrt{(2 \ell+1) (2 \ell'+1)}\, 
\left(\matrix{\ell & \ell' & 1 \cr 0 & 0 & 0}\right)\,
\langle \ell\, m\, \ell'\, -m' | 1\, -M_S \rangle\, \ \times \nonumber \\
&& \langle \frac{1}{2}\, \sigma\, \frac{1}{2}\, -\sigma' | 
1\, M_S \rangle\, 
\langle \ell  m  \frac{1}{2} \sigma  | j  \mu  \rangle\, 
b_{n  \ell  j   \mu }^{\dagger}\,
\langle \ell' m' \frac{1}{2} \sigma' | j' \mu' \rangle\, 
b_{n' \ell' j' \mu'}\, ,
\end{eqnarray}
which reduces into,
\begin{eqnarray}
D_{hh}(r)&=&\frac{1}{\sqrt{4 \pi}} \sum_{n n'\ell \ell' j} 
(-)^{j-\frac{1}{2}}\, 
\sqrt{(2 \ell+1) (2 \ell'+1)}\, 
\varphi_{n \ell}(r)\, \varphi_{n' \ell'}(r)
\nonumber \\
&&\left(\matrix{\ell & \ell' & 1 \cr 0 & 0 & 0}\right)\,
\left\{ \matrix{ l & l'& 1 \cr \frac{1}{2} & \frac{1}{2} & j} \right\}\, 
\sum_{\mu} b_{n  \ell  j   \mu }^{\dagger}\, b_{n' \ell' j \mu}\, ,
\label{Dhh}
\end{eqnarray}
where $\{\ \}$ is a Wigner $6j$ symbol. The equalities, $j=j'$ and $\mu=\mu',$
reflect the fact that the $LS$ coupling used in the previous section, Sec. 3, 
boils down to total spin $J=0,$ as demanded by the scalar nature of the 
many-body density operator ${\cal D}.$ Accordingly, in a $jj$ scheme, both 
the particle and the hole total spin labels must be equal.

The zeroth-order matrix element in $\varepsilon$ that results from 
Eqs. (\ref{pathol}) and (\ref{Dhh}), $\langle 0 | D_{hh} | 0 \rangle,$ 
trivially vanishes. Upon a simple inspection of the first-order matrix 
elements,
$\langle 0 |\, b_{n  \ell  j \mu }^{\dagger}\, b_{n' \ell' j \mu}\,
b_{\, 1s\frac{1}{2} \nu}^{\dagger}\, b_{\, 0p\frac{1}{2} \nu}\, | 0 \rangle,$
it is seen that the only non-vanishing contributions come from the cases, 
$\{n  \ell  j\mu\}=0p_{\frac{1}{2}}\nu$ and 
$\{n' \ell' j\mu\}=1s_{\frac{1}{2}}\nu$,
because of the restrictions on the values of $\ell$ and $\ell'$.
Here
$\nu$ denotes the magnetic label of both the particle and the hole in 
Eq. (\ref{pathol}).The two values of $\nu$ in  $|\tau \rangle$ give the 
same contribution, similarly to the two isospin components. With a global 
factor, $4\, \sqrt{3/\pi}\, \left(\matrix{1 & 0 & 1 \cr 0 & 0 & 0}\right)\,
\left\{ \matrix{ 1 & 0 & 1 \cr \frac{1}{2} & \frac{1}{2} & \frac{1}{2}} 
\right\},$ omitted, the $S=1$ form factor for the toy model reads,
\begin{equation}
D_{hh}(r) \propto \varphi_{01}(r)\, \varphi_{10}(r) = \pi^{-\frac{1}{2}}\,
\left( 2^{\frac{3}{2}} 3^{-\frac{1}{2}}\, r\, 
e^{-\frac{r^2}{2}}\right)\,
\left[6^{\frac{1}{2}} \left(1-\frac{2}{3} r^2\right)\,
e^{-\frac{r^2}{2}}\right].
\end{equation}

\begin{figure}
\mbox{ \epsfysize=45mm \epsffile{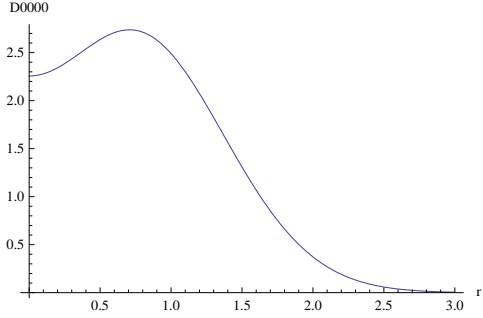} }
\caption{Scalar profile of the toy model}
\end{figure}

\begin{figure}
\mbox{ \epsfysize=45mm \epsffile{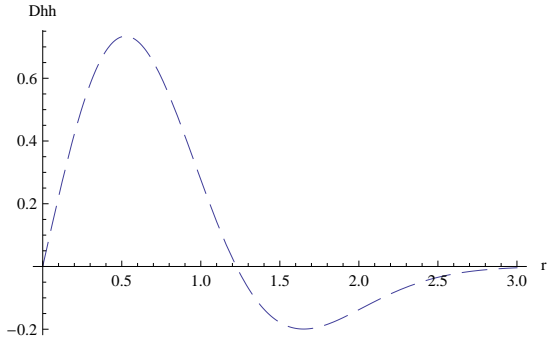} }
\caption{Vector profile of the toy model}
\end{figure}

Figures 1 and 2 show these scalar and vector profiles, respectively, in 
arbitrary units to avoid unessential global coefficients and because we 
prefer to compare shapes. There is no need to stress how different their 
shapes are.

\section{Discussion}
We set out to investigate the possible role of the spin density matrix in the 
construction of the density functional for nuclei. Such spin densities have 
played an important role in atomic and molecular physics. However, the severe 
constraints of rotational invariance and parity for nuclei led to the result
that the vector part of the spin density essentially vanishes in a nuclear DF 
that properly takes into account such symmetries, namely, in an RDF. Thus, 
there is no way, in this approach, to explicitly describe spin properties in 
a nuclear RDF. On the other hand, the vector part becomes a signature of 
parity violation allowed in the RDF theory. We were able to legitimize the
use of a spin density RDF, at the cost of introducing an external perturbation 
that has negative parity and transforms as a vector. Future studies are needed 
to understand the role of the spin-density-matrix formalism, when symmetries 
are broken by external forces.

\bigskip
{\it Acknowledgments}: B.R.B.and B.G.G. thank B.K. Jennings and T. 
Papenbrock for stimulating and helpful discussions.  
B.R.B. and B.G.G. also thank TRIUMF, Vancouver, B. C., Canada, 
for its hospitality, where part of this work was done. The Natural Science 
and Engineering Research Council of Canada is thanked for financial support. 
TRIUMF receives federal funding via a contribution agreement through the 
National Research Council of Canada.  B.R.B. also thanks Institut
de Physique Th\'eorique, CEA Saclay, France, for its hospitality, where
part of this work was carried out, and acknowledges partial support 
by NSF grants PHY-0555396 and PHY-0854912 and by Institut de
Physique Th\'eorique, CEA Saclay.

\end{document}